\documentclass[useAMS,usenatbib]{mn2e}
\usepackage{epsfig}
\usepackage{amsmath, amssymb,bm}

\title[Dead Zones as Sites for Satellite Formation]{\center Dead Zones In Circumplanetary Discs  as Formation Sites For Regular Satellites}

\author[ S. H. Lubow \& R. G. Martin] {Stephen
  H. Lubow and Rebecca G. Martin \\Space Telescope Science Institute, 3700 San Martin Drive,
 Baltimore, MD 21218, USA  }

\begin{document}

\date{}

\pagerange{\pageref{firstpage}--\pageref{lastpage}} 
\pubyear{2011}
\maketitle

\label{firstpage}

\begin{abstract}
Regular satellites in the solar system are thought to form within
circumplanetary discs. We consider a model of a layered circumplanetary
disc that consists of a nonturbulent midplane layer and and strongly
turbulent disc surface layers.  The dead zone provides a favorable
site for satellite formation. It is a quiescent environment that
permits the growth of solid bodies.  Viscous torques within the disc
cause it to expand to a substantial fraction of its Hill radius ($\sim
0.4 R_{\rm H}$) where tidal torques from the central star remove its
angular momentum.  For certain parameters, the dead zone develops into
a high density substructure well inside the Hill sphere. The radial
extent of the dead zone may explain the compactness of the regular
satellites orbits for Jupiter and Saturn.  The disc temperatures can
be low enough to be consistent with the high ice fractions of Ganymede
and Callisto.
\end{abstract}

\begin{keywords}
accretion, accretion discs -- planets and satellites: formation --
planetary systems
\end{keywords}

\section{Introduction}

After a planet forms in a circumstellar disc and before its mass
reaches a value of Jupiter's mass, tidal forces from the planet open a
gap in the disc \citep{lin84,lin86,dangelo02,bate03}. The planet
continues to accrete material from the circumstellar disc through the
gap and a circumplanetary disc forms \citep{artymowicz96,
  lubow99,dangelo03, ayliffe09}.  A circumplanetary (spin-out) disc
may also form if the planet manages to achieve break-up velocities as
it contracts \citep{korycansky91}. The regular satellites in the solar
system have prograde, nearly circular, and nearly coplanar orbits.
They are thought to have formed in a circumplanery disc, in analogy
with the formation of planets in the solar nebula \citep{pollack91}.
Understanding the structure of these discs is essential for explaining
regular satellite formation.

There are several key constraints that various models of gaseous
circumplanetary discs have attempted to satisfy. The regular
satellites around Jupiter and Saturn lie within a radius of less than
$0.06 R_{\rm H}$ of their Hill (tidal) radii $R_{\rm H}$.  The small
extent of this region has been explained to be a consequence of a
compact disc structure or substructure.  Various studies have tied the compactness to
the level of angular momentum of accreting material entering a planet's
Hill sphere \citep[e.g.,][]{estrada09, ward10}. The idea is that if the
accreting inflow has sufficiently low angular momentum per unit mass,
then the satellites formed with a circumplanetary disc will occupy a small
region within the Hill sphere.

All hydrodynamics simulations of gas flows about a Jupiter mass planet
have revealed a disc that is considerably more extended than the
Gallilean satellites.
 The recent study by \cite{ayliffe09} shows that
the disc extends to about $0.35 R_{\rm H}$. For a planet such as
Jupiter that opens a gap in the circumstellar disc, gas enters the
Hill sphere preferentially near the Lagrange points and carries
significant angular momentum.   However, the main issue is that the
compactness of a fully turbulent disc cannot be maintained due to the
action of viscous torques \citep{ward10, martin11a}. These torques
cause a compact circum-Jovian disc to rapidly expand in only $\sim
100$ planet orbital periods for typical disc parameters.  

Although
the angular momentum of the inflowing
gas is  uncertain, the radial extent of the disc is largely independent
of this quantity \citep{martin11a}. The disc
expands to a radius where the tidal torques from the central star
(Sun) remove angular momentum at the rate it is supplied by inflowing
gas. This radius is estimated as about $0.4 R_{\rm H}$, where
ballistic periodic orbits begin to cross \citep{martin11a}. (Some
departures from this value occur due to pressure effects.)  In
addition, no compact substructure is found to lie within such a
circumplanetary disc \citep{estrada09, ayliffe09, martin11a}.  With
this simple alpha disc model, the turbulent viscosity is by assumption
a smooth function of radius.  The disc structure then follows that of
a standard smooth accretion disc.

Previous work such as \cite{canup02} 
suggested that if the inflowing angular momentum is small enough,
 the flow may penetrate deep into the Hill sphere before initially collecting into a disc.  
If the disk is viscous, its gas component will then spread outward to a large radius.  
But the solids initially delivered with the gas may not be effectively coupled to the gas if they rapidly accrete and grow once in circumplanetary orbit.  
The solids then accrete near the compact region where they were initially delivered, rather than tracking the outward viscous expansion of the gas disk.  

There are some issues of concern with this picture. 
The inflow may be largely planar and join the circumplanetary disk at its outer edge,
as indicated by simulations by Ayliffe \& Bate (2009). The gas and coupled
solids would
then not be able to penetrate deeply within the Hill sphere. Instead, they achieve Keplerian orbits near the outer edge
of the pre-existing disk as they becomes entrained there.  It is possible that larger size solid bodies
that are decoupled from the gas could enter the Hill sphere with sufficiently low angular
momentum to be captured close to the planet. 
 But it is not clear that such a low
angular momentum occurs.
In this paper, we suggest an alternative
explanation for the compactness of the satellites based on the radial extent
of a dead zone, a region of low disc turbulence.

Another constraint applies to the disc temperature.  This snow line
plays an important role in the composition of forming
satellites. Outside of the snow line, the solid mass density is much
higher because of water ice condensation.  The snow line occurs at a
temperature, $T_{\rm snow}$, that is around $170\,\rm K$
\citep{hayashi81,lecar06}. Ganymede and Callisto, the two outer
Gallilean satellites contain a substantial fraction of ice ($\sim
50\%$), while the two inner satellites contain much less. The snow line
of the disc is then required to have been near the orbits of these
outer satellites.  The partially differentiated structure of Callisto
suggests that its ice never fully melted and that the snow line was
always inside its orbit \citep{lunine82}. Since the disc temperature
increases with mass accretion rate, this constraint limits that rate.
This requirement implies that the disc accretion rates be sufficient
low, substantially lower than what would be expected during the T
Tauri accretion phase \citep{canup02, estrada09}. The satellite
formation epoch is therefore expected to occur late in the life of the
solar nebula, when it has lost a substantial fraction of its mass.

Another issue involved in the formation of both planets and satellites
in gaseous discs is the permitted level of turbulence.  Estimates of
disc turbulence from simulations and properties of observed systems
suggest that relatively high levels of turbulence $\alpha \sim 0.01 -
0.3$ are expected in fully turbulent discs \citep[e.g.,][]{king07}.
Larger solid bodies may form from the gravitational instability of a
dense layer of dust that settles near the disc mid-plane
\citep[e.g.,][]{goldreich73}.  Even modest levels of turbulence,
involving $\alpha \ll 0.01$, can prevent solids from setting to a thin
enough layer near disc mid-plane for gravitational instability to
operate \citep{dubrulle95, cuzzi06}.  An alternative model for the
growth of solids involves the concentration of dust in turbulent
eddies \citep{cuzzi08}. This model also adopted a weaker level of
turbulence. Another effect is that turbulence can cause destructive
collisions among larger solid bodies, thereby preventing growth to the
desired size \citep[e.g.,][]{Ida08}.  These results suggest that lower
values of disc turbulence may be more favorable for satellite
formation than is occurs in a fully turbulent disc.

In a layered disc, the magnetic turbulence \citep[MRI;][]{balbus91} is
not active at all heights \citep{gammie96}. Magnetic turbulence
requires a certain level of ionization for the gas to be well enough
coupled to the magnetic field. In sufficiently cool disc regions,
temperatures are too low to provide the needed level of the ionization
for magnetic turbulence to operate.  Instead, the ionization is
provided by external sources of radiation, such as X-rays or cosmic
rays. If the disc surface density is high enough, this radiation may
not penetrate deep enough below the disc surfaces to provide
turbulence at all heights.  A so-called dead zone with little or no
turbulence develops around the disc midplane. Such an environment may
be favorable for the survival and growth of solid bodies.

In an earlier paper, we explored a layered disc model for the
evolution of a circumplanetary disc during a stage when the disc
experiences accretion rates comparable to those of the T Tauri phase
\citep{lubow12}.  Such discs can sometimes undergo outbursts,
analogous to the FU Ori outbursts.  In any case, such a disc is too
hot for the survival of icy satellites at the locations of Ganymede
and Callisto.

We explore in this paper the viability of a layered disc model as an
environment for satellite formation in later stages of its
evolution. We do not attempt to model the formation of satellites in
such a disc.  Instead, we show that in this stage and for certain
plausible disc ionization parameters, a layered disc model satisfies
the requirements described above.  In particular, a compact high
density disc substructure develops in a low turbulence dead zone whose
extent matches that of the regular satellites. The disc temperature is
low enough for the outermost Galilean satellites to lie outside the
snow line.

In Section~\ref{runs}, we describe the results from our
circumplanetary disc model and how it may explain some features of the
formation of the Galilean satellites. In Section~\ref{dis} we discuss
some implications and limitations of the model and in
Section~\ref{conc} we draw our conclusions.

\section{Circumplanetary Disc Model}
\label{runs}

In this Section we first describe our layered circumplanetary disc
model that is truncated by tides from the star and then we discuss the
results.

\subsection{Model Description}

We use the layered disc model initially described in \cite{armitage01}
and further developed in \cite{zhu09}, \cite{martin11b} and
\cite{lubow12}.  Material in the circumplanetary disc orbits the
central planet of mass $M_{\rm p}$ at Keplerian angular speed
$\Omega(R)=\sqrt{G M_{\rm p}/R^3}$ for radius $R$ from the planet.
Material is added to the disc at rate $\dot M_{\rm infall}$. The disc
has a total surface density $\Sigma(R,t)$, midplane temperature
$T_{\rm c}(R,t)$, and surface temperature $T_{\rm e}(R,t)$ at time
$t$. The MRI turbulent surface layer (which we call the active layer)
has surface density $\Sigma_{\rm m}(R,t)$, temperature $T_{\rm
  m}(R,t)$, and turbulent viscosity $\nu_{\rm m}(T_{\rm m},R,t)$ that
is parametrised with the \cite{shakura73} $\alpha_{\rm m}$ parameter.
Cosmic rays and/or X-rays are assumed to be able to provide sufficient
ionization for MRI to operate to a constant surface density depth
$\Sigma_{\rm crit}$. If the total surface density is greater than this
critical value, $\Sigma>\Sigma_{\rm crit}$, then $\Sigma_{\rm
  m}=\Sigma_{\rm crit}$.  For smaller surface density,
$\Sigma<\Sigma_{\rm crit}$, the disc is fully MRI active and
$\Sigma_{\rm m}=\Sigma$.

A complementary midplane layer exists if the total surface density is
larger than the critical, $\Sigma>\Sigma_{\rm crit}$. It has surface
density $\Sigma_{\rm g}=\Sigma-\Sigma_{\rm crit}$ and midplane
temperature $T_{\rm c}$. This layer is MRI active if the midplane
temperature is greater than the critical, $T_{\rm c}>T_{\rm crit}$,
where $T_{\rm crit}$ is the temperature for sufficient thermal
ionisation.  However, for lower temperatures, $T_{\rm c}<T_{\rm
  crit}$, a dead zone forms.  In the model, there is no turbulence in
the dead zone unless it is massive enough to be self-gravitating.  (In
reality, some turbulence is expected due to the vertical propagation
of waves from the active layer \citep{fleming03}.) The condition for
self gravity is taken to be that the Toomre parameter is smaller than
its critical value, $Q<Q_{\rm crit}=2$ \citep{toomre}. In this case, a
second viscosity term is included in the complementary layer.

We consider the planet Jupiter with mass, $M_{\rm p}=1\,M_{\rm J}$,
that is orbiting the Sun, $M=1\,\rm M_\odot$, at a distance of
$a=5.2\,\rm AU$. We solve the accretion disc equations numerically on
a fixed mesh that is uniform in $\log R$ with 120 grid points
\citep[e.g.,][]{armitage01,martin07}. The inner boundary has a zero
torque condition at the radius of Jupiter, $R=1\,R_{\rm J}$, so the
mass falls freely on to the planet. At the outer boundary we
approximate the tidal torque from the star with a zero radial velocity
boundary condition at $R_{\rm out}=0.4\,R_{\rm H}$.  Material accretes
on to the disc at a constant rate $\dot M_{\rm infall}$ at a radius
which we take to be $0.33\, R_{\rm H}$ \citep[see][for more
  explanation]{lubow12}. Initially, we take the surface density and
temperature structure of the disc to be that of low mass (nonself
gravitating) disc with a small dead zone. This state represents the
disc immediately following an outburst. Alternatively, if there is
initially no mass in the disc, it quickly builds up to the same state.

The critical surface density that is sufficiently ionised to be
turbulent, $\Sigma_{\rm crit}$, is not well determined. When cosmic
rays are the dominant source of ionisation, for typical parameters the
surface turbulent layer of a circumstellar disc has been estimated to
have a surface density of $\Sigma_{\rm crit}\approx 200\,\rm
g\,cm^{-2}$ \citep{gammie96,fromang02}. However, it is possible that
cosmic rays are swept away from the disc by means of a turbulent MHD
jet or outflow which may be present at the star
\citep[e.g.][]{skilling76,cesarsky78}. In the absence of cosmic rays,
X-rays from the star may be the dominant source of ionisation and in
this case the active layer is much smaller \citep{matsumura03}.

More recent works find the inclusion of polycyclic aromatic
hydrocarbons and dust tends to suppress the instability further
\citep{bai09, perez11,bai11, fujii11}. However, in circumstellar
discs, such effects produce an accretion rate that is too low to
account for T Tauri accretion rates and this problem remains
unsolved. \cite{perez11} suggest that in order to explain the T Tauri
rates, $\Sigma_{\rm crit}>10\,\rm g\,cm^{-2}$, in the circumstellar
disc. It is possible that circumplanetary discs may have an even
smaller active layer surface density than circumstellar discs, since
they may be shielded from the ionising radiation of the central star
by the shadowing effects of the inner circumstellar disc. Because of
these uncertainties, we follow the work of Armitage et al (2001) and
\cite{zhu09} and regard $\Sigma_{\rm crit}$ as a constant free
parameter. If the critical surface density is large, then the entire
disc will be fully turbulent. If it is small, then the dead zone will
be very extensive.  We consider a range of active layer surface
densities with $1\, \rm g\,cm^{-2} \le \Sigma_{\rm crit} \le 100 \,
\rm g\,cm^{-2}$.

Generally, we take the value of the temperature for sufficient thermal
ionization for MRI to operate to be $T_{\rm crit}=800\,\rm K$. At this
temperature the ionisation fraction increases exponentially with
temperature due to the collisional ionisation of potassium
\citep{umebayashi83}. The viscosity $\alpha_{\rm m}$ parameter
associated with MRI is found to be $\gtrsim 0.01$ from MHD
simulations, but depends on numerous parameters such as the
resolution, stratification, and treatments of small scale dissipation
and radiation transport \citep[e.g.][]{hartmann98,fromang07,guan09,
  davis09}. Similarly, observations of FU Ori suggest that
$\alpha_{\rm m}\approx 0.01$ \citep{zhu07}. However, observations of
X-ray binaries and dwarf novae suggest $\alpha_{\rm m} \approx
0.1-0.4$ \citep{king07}.  Armitage et al (2001) take $\alpha_{\rm m}=0.01$
and \cite{zhu10} consider both $0.01$ and $0.1$. With the uncertainty
in the value, we also consider cases with both $\alpha_{\rm m}=0.01$
and $0.1$.

During times of planet formation, the accretion rate on to the
circumplanetary disc, $\dot M_{\rm infall}$, is of order the overall
circumstellar disc accretion rate \citep{bate03,lubow06,ayliffe09}
that is of order $10^{-8}\,\rm M_\odot\,yr^{-1}$
\citep{valenti93,hartmann98}, typical for the T Tauri phase. During
this phase, disc temperatures are too high to permit the survival of
ice in Callisto \citep{canup02, estrada09}.  Consequently, we consider
the evolution in the later stages of disc evolution, when the
accretion rate has dropped by an order of magnitude or more. We
consider accretion rates in the range $10^{-11} - 10^{-9}\,\rm M_\odot
\, yr^{-1}$ near the end of the disc lifetime.

If the circumplanetary disc density in the dead zone grows sufficient large during the T Tauri stage, the disc
becomes gravitationally unstable. Self-gravity generates turbulence
that raises the disc midplane temperature
\citep[e.g.,][]{lodato04}. If the temperature exceeds the critical
value $T_{\rm crit}$ for sufficient thermal ionization for MRI to
operate, MRI can set in abruptly and lead to a gravo-magneto accretion
outburst in a circumplanetary disc \citep{lubow12}.  These
conditions involve high disc temperatures. The dead zone structure is
disrupted and dead zone mass is accreted on the central planet.  The
process repeats and can be described as a limit cycle of the
gravo-magneto instability \citep{martin11b}.  Consequently, we regard
the T Tauri stage as unfavorable for satellite formation or survival.

As the disc disperses near the end of its lifetime, the accretion rate
$\dot M_{\rm infall}$ decreases and outbursts become less likely.  We
are most interested in models that have a long outburst interval
timescale. The outburst interval timescale increases with decreasing
accretion rate. If the mass accretion rate decreases substantially
after an outburst, then the outburst may not recur.  The lifetime of
the circumplanetary disc is assumed to be similar to that of the
circumstellar disc, a few $10^6\,\rm yr$.  A simple estimate for the
outburst timescale is the timescale for the disc to become self
gravitating. This timescale can be estimated as the time to build up a
mass of $M_{\rm p}H/R$, where $H=c_{\rm s}/\Omega$ is the disc scale
height and $c_{\rm s}$ is the sound speed.  This mass is at most
around a few tenths of a Jupiter mass. The outburst interval timescale
is of order
\begin{equation}
t_{\rm int} \approx \frac{M_{\rm p}}{\dot M_{\rm
  infall}}\frac{H}{R}.
\end{equation}

\subsection{Model Results}

\begin{figure*}
\includegraphics[width=8.4cm]{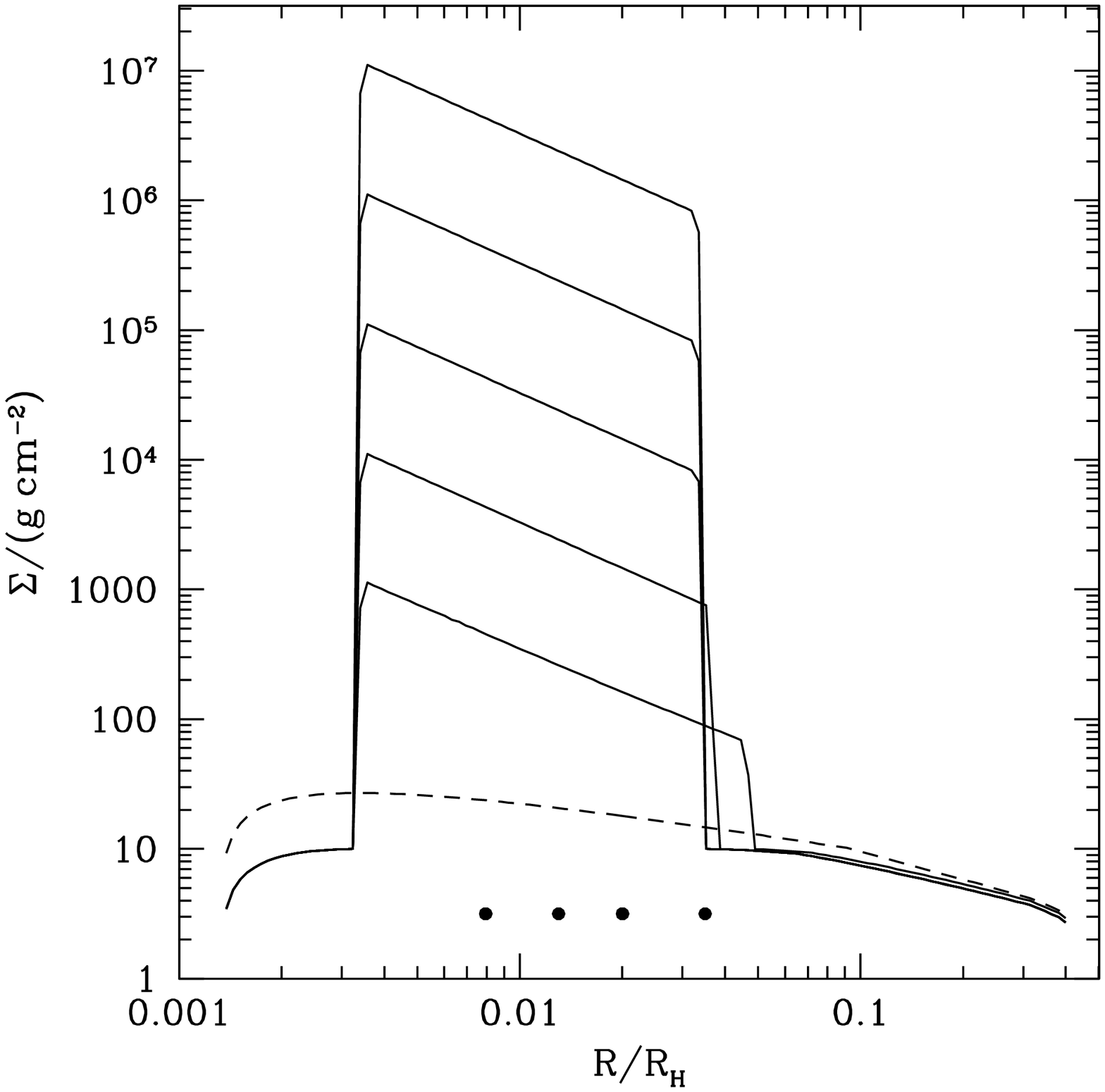}
\includegraphics[width=8.4cm]{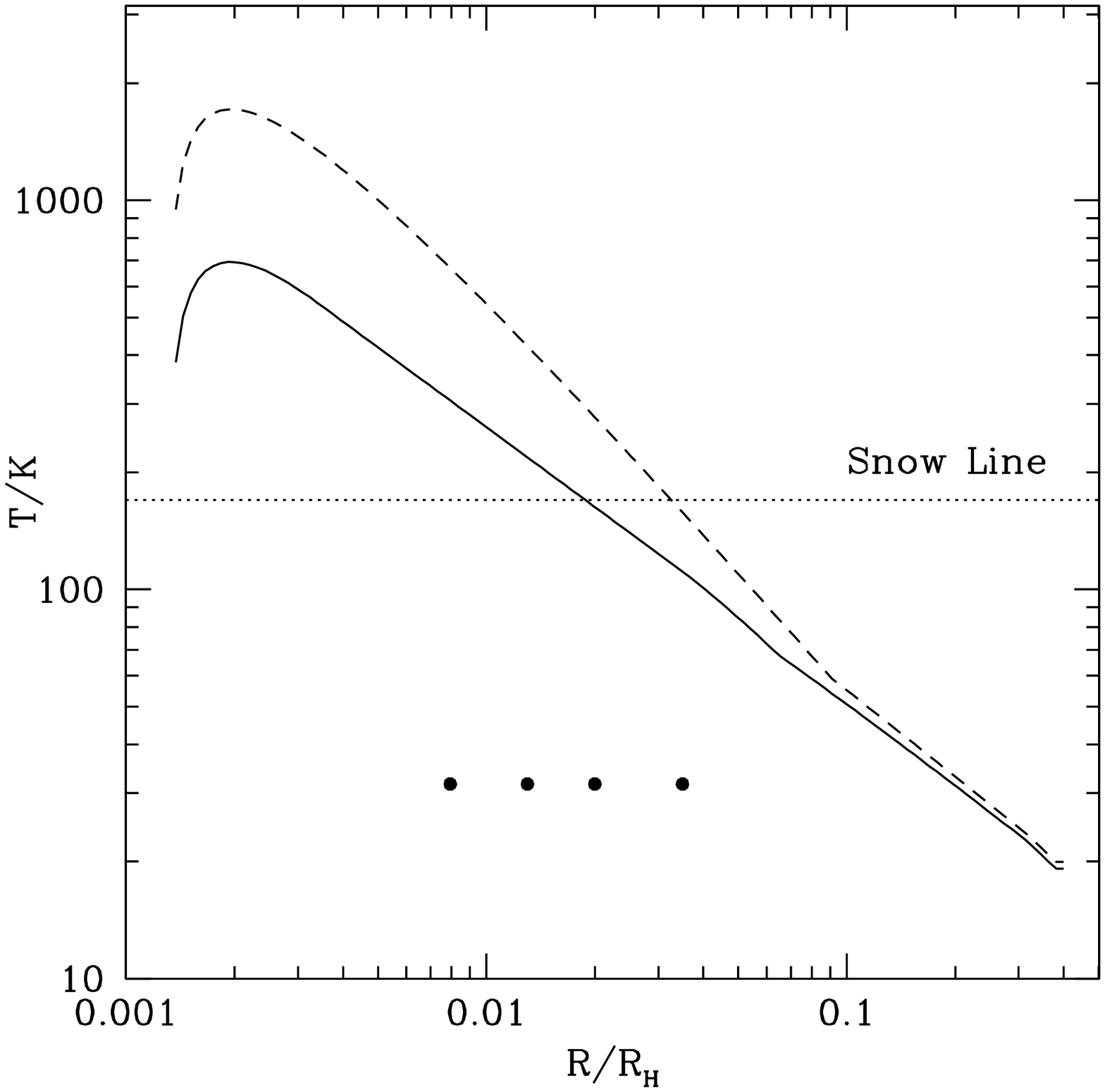}
\caption{Left: The surface density of the disc in model R8 (dashed
  line) and R9 (solid lines) at times $t=10^2$, $10^3$, $10^4$, $10^5$
  and $10^6\,\rm yr$ in order of increasing height in the plot.
  As the dead zone gains mass, its surface density grows in time.  
  Right: The midplane temperature of the
  disc in R9 (solid line) and R8 (dashed line). The temperature in
  model R9 does not change in time even though the surface density
  does.  The dotted line shows the snow line temperature, $T_{\rm
    snow}=170\,\rm K$. The solid circles in both plots show the radial
  locations of the Galilean satellites.}
\label{surf}
\end{figure*}

 Table~\ref{table} summarizes simulation results for a range of
 parameters that cover some plausible values for the viscosity
 parameter $\alpha_{\rm m}$ and the critical surface density for
 ionization $\Sigma_{\rm crit}$.  In Fig.~\ref{surf} we show the
 surface density and temperature distributions for models with
 $\alpha_{\rm m}=0.01$ and $\dot M_{\rm infall}=10^{-10}\,\rm
 M_\odot\,yr^{-1}$. Model R8 is fully turbulent and this steady
 solution holds for $\Sigma_{\rm crit}>27.1\,\rm g\,cm^{-2}$.  Model
 R9 has critical surface density $\Sigma_{\rm crit}=10\,\rm
 g\,cm^{-2}$ and hence contains a dead zone.  Within the dead zone,
 the surface density increases substantially with time.  For Model R9
 this region extends to radius $R_{\rm dead}=0.034\,R_{\rm H}$. This
 location is close to the orbit of Callisto. Such a model may explain
 why Callisto has a large mass and angular momentum, yet no satellites
 are found outside of its orbit. Beyond the dead zone, less material
 is available to form satellites and the turbulence may disrupt the
 growth into larger bodies.

At later times, the higher disc mass in the dead zone is in the range
of the so-called minimum mass sub-nebula (MMSN)  \citep{lunine82}.
However, the model here is fundamentally different from a static MMSN that was
initially envisioned by some earlier work, because it considers a continuously supplied disc.
The accretion time scale in this paper is regulated by the inflow rate to the disc, as in the original continuous inflow concept of \citep{canup02}.  

The disc
structure we obtain is similar to the gas rich disc model envisioned
by \cite{mosqueira03a} and \cite{estrada09}.  In their model, Callisto
lies just outside the dense inner disc, the dead zone in the current
model.  They suggested that Callisto's inferred long formation
timescale \citep{stevenson86}, which caused its partial
differentiation, results from the relatively slow delivery of solids
in the outer disc, due to its lower density. The model also describes
Ganymede as being formed more rapidly within the dense disc.  These
different environments provide a possible explanation for the
Callisto-Ganymede dichotomy that describes the differences between
these two outermost Gallilean satellites \citep{lunine82}.

The mass of the fully turbulent disc in Model R8 is $4.3\times
10^{-5}\,\rm M_{\rm J}$ whereas the mass of the disc with a dead zone
in Model R9 at a time of $10^5\,\rm yr$ is $9.1\times 10^{-3}\,\rm
M_{\rm J}$. The dead zone model has a lower temperature distribution
than the fully turbulent model. For example, the snow line location
changes from $0.032\,R_{\rm H}$ in the fully turbulent model to
$0.017\,R_{\rm H}$ when a dead zone is introduced.  The latter snow
line location is in the range expected for the survival of ice in
Callisto.  The outburst intervals for Model R9 at constant accretion
rate are also fairly long.  For model R9, $H/R<0.2$ and so the
outburst timescale is around $2\times 10^6\,\rm yr$.

With a higher accretion rate of $10^{-9}\,\rm M_\odot\,yr^{-1}$, the
disc builds up mass very quickly and so the outburst time scale is
fairly short. In addition, it is difficult to simultaneously meet the
constraints on the snow line location and the compactness of the dead
zone. The models that best meet these requirements, Models R9 and R12,
have the intermediate accretion rate $\dot M=10^{-10}\,\rm
M_\odot\,yr^{-1}$.

We note that because of the dead zone, the vertically averaged
$\alpha$ in the disc is much lower than the value in the active layer
of $0.01$ or $0.1$.  In some ways this vertical average approximates
the $\alpha$ value adopted by \cite{canup02}. However, since that
model does not have a dead zone, there is no accumulation of mass
that produces the compact disc substructure.
Fully turbulent models with lower accretion rates of $10^{-11}\,\rm
M_\odot\,yr^{-1}$ (R13 and R14) have low enough temperatures for icy
satellite formation \citep[as found by][]{canup02}.

\begin{table*}
\centering
\begin{tabular}{ccccccccccccc}
\hline
\hline
Model   &  $\dot M_{\rm infall}$ &   $\alpha_{\rm m}$ & $\Sigma_{\rm crit}$ & $R_{\rm dead}$ & $R_{\rm snow}$  & $M_{\rm disc}(t=10^5\,\rm yr)$ & {\rm Outburst Timescale}\\
    &     $(\rm M_\odot\,yr^{-1})$  &    &  $\rm (g\,cm^{-2})$  &  ($R_{\rm H}$) &  ($R_{\rm H}$) & ($M_{\rm J}$)  & (yr) \\
\hline
\hline
R1        &      $10^{-9}$   &    0.01  &   $>90.3$   & -          &  0.094  & $1.9\times 10^{-4}$ & {\rm Steady}  \\
R2      &      $10^{-9}$   &    0.01  &   40        &  0.023     &  0.084  & $6.7\times 10^{-2}$  &   $1.4\times 10^5$ \\
R3     &      $10^{-9}$   &    0.01  &   10        &  0.32      &  0.019  & $1.0\times 10^{-1}$  &  $7.8\times 10^5$ \\
R4     &      $10^{-9}$   &    0.01  &   1         &  0.33      &  0.0023 & $1.1\times 10^{-1}$  &   $8.3 \times 10^5$ \\
R5    &      $10^{-9}$   &    0.1   &   $>15.6$   & -          &  0.056  & $2.5\times 10^{-4}$   &   {\rm Steady}\\
R6     &      $10^{-9}$   &    0.1   &   10        &  0.0097    &  0.052  &         &   $> 10^5$\\
R7     &      $10^{-9}$   &    0.1   &   1         &  0.32      &  0.011  & $1.0\times 10^{-1}$  &   $> 10^5$\\
\hline
R8      &      $10^{-10}$  &    0.01  &   $>27.1$   &   -        &  0.032  & $4.3\times 10^{-5}$  &   {\rm Steady}  \\
\bf{R9} &      $\bf{10^{-10}}$  &  \bf{0.01}  &   \bf{10}        &  \bf{0.034}  &  \bf{0.017}  & $\bf 9.1\times 10^{-3}$  &   $\bf > 10^6$ \\
R10     &      $10^{-10}$  &    0.01  &   1         &  0.32      &  0.0023 & $1.1\times 10^{-2}$  &   $> 10^6$ \\
R11       &      $10^{-10}$  &    0.1   &   $>4.7$    &   -        &  0.021  & $4.5\times 10^{-6}$  &   {\rm Steady}  \\
\bf{R12} &      $\bf 10^{-10}$  &    \bf 0.1   &  \bf  1         & \bf 0.035     & \bf 0.011  & $\bf 9.8\times 10^{-3}$  &   $\bf > 10^6$ \\
\hline
R13  	&      $10^{-11}$  &    0.01  &   $>8.1$   & -   &  0.011  & $7.7\times 10^{-6}$ &   {\rm Steady}\\
R14     &      $10^{-11}$  &    0.1   &   $>1.2$   & -   &  0.0092 & $7.8\times 10^{-7}$ &  {\rm Steady} \\
\hline
\end{tabular}
\caption{ Column~2 contains the infall accretion rate on to the
  circumplanetary disc, column~3 contains the viscosity $\alpha_{\rm m}$
  parameter, column~4 contains the critical surface density in the active
  layer that is ionised by cosmic rays/X-rays, column~5 contains the
  outer radius of the dead zone, $R_{\rm dead}$, if it exists,
  column~6 contains the snow line radius and column~7 contains the total
  mass in the disc at time $t=10^5\,\rm yr$. If the disc has a dead
  zone, then the mass of the disc increases linearly in time. But if
  there is no dead zone, then the mass in constant. Finally, column~8
  contains estimates of the timescales for the gravo-magneto outbursts. }
\label{table}
\end{table*} 

\section{Discussion}
\label{dis}

There are several approximations in the layered disc model we have
considered that should be improved on in future work. Some of these
were discussed in \cite{martin11b} and \cite{lubow12}. For example, we
have approximated the surface density in the active layer as a
constant with radius in the disc.  An alternative method of finding
the dead zone is with a critical magnetic Reynolds number
\citep[e.g.][]{fromang02,matsumura03}. In this approach, the surface
density of the turbulent layer may vary in radius \citep{martin12a}.
However, there are currently still problems with this model, as it is
difficult to explain T Tauri accretion rates with a low critical
surface density \citep{martin12b}.

The infall accretion rates we considered are constant in time.
However, they are likely to decrease as the flow through the
circumstellar disc decreases. As the accretion rate decreases, the
active surface layer occupies an increasing fraction of the vertical
extent of the disc. When the accretion rate drops to values of
$\la10^{-11} \rm M_\odot\,yr^{-1}$, the disc will become fully
turbulent with the adopted parameters (Models R13 and R14).  The disc
temperatures remain low enough for the survival of ice in Callisto.

The inferred density of the small inner Jovian satellite Amalthea
suggests that contains a substantial fraction of water-ice
\citep{anderson05}.  Accretion disc models, such as the one described
here, predict that temperatures are too high for the water-ice to
survive and requires some other explanation.

Another open issue is satellite survival in such a high density disc
against the effects of orbit decay due to disc-satellite interactions
\citep{ward97, lubow10}.  The Type I migration timescale in the dead
zone of model R9 is quite short $\sim 10^2 \rm yr$ \citep{canup02}.
It is possible that the satellites instead undergo the potentially
slower Type II migration, provided that they can open gaps.  Gap
opening in a low viscosity disc is dominated by effects of gas shocks
caused by the steepening of density waves \citep{rafikov02, yu10}.
The gap opening condition in this case suggests that Callisto and
Ganymede can open gaps for Model R9, but only for $\Sigma \la 10^4 \rm
g\, cm^{-2}$.

The nature of gap opening and migration in a layered disc has not been
examined. The application of the viscous gap opening criterion
\citep[e.g., equation 5 of][]{canup02} to the active layer alone
suggests that gap opening does not occur. On the other hand, its
application to a vertical averaged $\alpha$ suggests that gap opening
may occur. There are some other possibilities. If Callisto resides
just outside the dead zone, it may be subject to outward Lindblad
torques that stall its migration there \citep{mosqueira03b,
  matsumura07}.  Trapping may also be possible at the inner edge of
the dead zone \citep{kretke09}. We also note that outward corotation
torques could affect migration \citep{paardekooper06}, since the dead
zone region in Fig.~1 has a negative radial entropy gradient that is
required for this effect to operate.  On the other hand, coorotation
torques can become quite weak and saturate in a low viscosity
environment \citep[e.g.,][]{ward91, ogilvie03}.

\cite{canup02} point out other constraints that should be considered.
One is that if satellites open gaps in the disc, then satellite
eccentricities should be excited by first order Lindblad resonances.
The excitation of eccentricity of planets that open gaps has been an
active area of exploration \citep{goldreich03, ogilvie03}. The
simulation results appear to be sensitive to how clean and large the
gap is.  \cite{pap01} found no eccentriciy growth for a $1 M_{\rm J}$
which produces a fairly clean gap, but growth at much higher
masses. \cite{dangelo06} found some eccentricity growth occurred $1
M_{\rm J}$, but was quite weak compared to growth rates for higher
mass planets.  In any case, at late times when $\dot M_{\rm infall}
\la 10^{-11} \rm M_\odot\,yr^{-1}$, the disc surface density in the
satellite region will drop to less than critical values $\Sigma(R) <
\Sigma_{\rm crit}$. The disc is then fully turbulent and gap opening
will not occur.  The available gas at this stage may be sufficient to
damp eccentricities developed in a prior open-gap stage.

The disc structure produced by this model is somewhat similar to the
gas rich model described in \cite{mosqueira03a} and
\cite{estrada09}. In that model, the dense inner disc is the relic of
a disc formed prior to gap opening.  The low density outer disc is
formed by higher angular momentum gas accreted after gap opening.  It
is not clear why the inner disc would not accrete onto the planet. We
expect that a significant fraction of the mass of Jupiter was acquired
after gap opening through disc accretion.  In the present model,
accreting gas builds up mass in a small inner region due to the
presence of the dead zone.

The dead zone is assumed to be free of turbulence.
Suppose some independent mechanism produces turbulence there
that we do not take into account, such as a hydrodynamic instability
not involving MRI or self-gravity. The dead zone would 
then achieve a steady state flow that would limit the mass growth
of the dead zone as seen in Fig.~\ref{surf}. 
For the parameters adopted in Model R9,  reaching the highest plotted densities
requires that $\alpha_{\rm d} \sim 10^{-8}$.

\section{Conclusions}
\label{conc}

We have described some possible proto-satellite environments that
occur in discs with dead zones.  A dead zone will occur if external
radiation is insufficient to fully ionize cool regions of a disc
\citep{gammie96}.  The dead zone provides a quiescent environment for
the survival and growth of solid bodies into satellites.  As the dead
zone gains mass, it can provide a high density, compact substructure
within the circumplanetary disc (see Fig.~1).  The regular satellites
of Jupiter and Saturn lie within a small fraction of their Hill radii.
A fully turbulent simple alpha disc extends to much larger radii and
has a smoothly varying density structure.  The compactness of a high
density dead zone provides a possible explanation for the small radial
extent of the regular satellites.  For accretion rates appropriate to
the late stages its evolution, the circum-Jovian disc can be cool
enough to permit the survival of the ice in Callisto.


In this paper we have only considered possible disc structures that
can arise in circumplanetary discs with dead zones.  More work is
required to explore the growth, survival, and evolution of satellites
in such discs.

\section*{Acknowledgements}

SHL acknowledges support from NASA grant NNX11AK61G.
RGM thanks the Space Telescope Science Institute for a Giacconi
Fellowship.  We thank  the referee for helpful comments.


\label{lastpage}
\end{document}